\documentclass[twocolumn, unsortedaddress, superscriptaddress]{revtex4}
\usepackage{graphicx}
\usepackage{color}
\usepackage{amsmath}

\newcommand{\be}{\begin{equation}}
\newcommand{\ee}{\end{equation}}
\newcommand{\bea}{\begin{eqnarray}}
\newcommand{\eea}{\end{eqnarray}}

\begin{document}

\title{Electrical control of optical emitter relaxation pathways enabled by graphene}

\author{K.J. Tielrooij}  \affiliation{ICFO - Institut de
Ci\'encies Fot\'oniques, Mediterranean Technology Park,
Castelldefels (Barcelona) 08860, Spain}
\author{L. Orona}\thanks{Equal contribution} \affiliation{Department of Physics, Massachusetts Institute of Technology, Cambridge, MA 02139, USA}
\author{A. Ferrier}\thanks{Equal contribution}\affiliation{Institut de Recherche de Chimie Paris, CNRS-Chimie Paristech, 11 rue Pierre et Marie Curie, 75005, Paris, France} \affiliation{Sorbonne Universit\'es, UPMC Univ Paris 06, 75005, Paris, France}
\author{M. Badioli}\thanks{Equal contribution}\affiliation{ICFO - Institut de Ci\'encies Fot\'oniques,
Mediterranean Technology Park, Castelldefels (Barcelona) 08860,
Spain}
\author{G. Navickaite}\affiliation{ICFO - Institut de Ci\'encies Fot\'oniques,
Mediterranean Technology Park, Castelldefels (Barcelona) 08860,
Spain}
\author{S. Coop}\affiliation{ICFO - Institut de Ci\'encies Fot\'oniques,
Mediterranean Technology Park, Castelldefels (Barcelona) 08860,
Spain}
\author{S. Nanot}\affiliation{ICFO - Institut de Ci\'encies Fot\'oniques,
Mediterranean Technology Park, Castelldefels (Barcelona) 08860,
Spain}
\author{B. Kalinic}\affiliation{Physics and Astronomy Department and CNISM, University of Padova,
via Marzolo 8, I-35131 Padova, Italy}
\author{T. Cesca}\affiliation{Physics and Astronomy Department and CNISM, University of Padova,
via Marzolo 8, I-35131 Padova, Italy}
\author{L. Gaudreau}\affiliation{ICFO - Institut de Ci\'encies Fot\'oniques,
Mediterranean Technology Park, Castelldefels (Barcelona) 08860,
Spain}
\author{Q. Ma}\affiliation{Department of Physics, Massachusetts Institute of Technology, Cambridge, MA 02139, USA}
\author{A. Centeno}\affiliation{Graphenea SA, 20018 Donostia-San Sebasti\'ian, Spain}
\author{A. Pesquera}\affiliation{Graphenea SA, 20018 Donostia-San Sebasti\'ian, Spain}
\author{A. Zurutuza}\affiliation{Graphenea SA, 20018 Donostia-San Sebasti\'ian, Spain}
\author{H. de Riedmatten}\affiliation{ICFO - Institut de Ci\'encies Fot\'oniques,
Mediterranean Technology Park, Castelldefels (Barcelona) 08860,
Spain} \affiliation{ICREA-Instituci\'o
Catalana de Recerca i Estudis Avan\c{c}ats, Passeig Llu\'{\i}s Companys,
23, 08010 Barcelona, Spain}
\author{P. Goldner}\affiliation{Institut de Recherche de Chimie Paris, CNRS-Chimie Paristech, 11 rue Pierre et Marie Curie, 75005, Paris, France}
\author{F.J. Garc\'ia de Abajo}\affiliation{ICFO - Institut de Ci\'encies Fot\'oniques,
Mediterranean Technology Park, Castelldefels (Barcelona) 08860,
Spain} \affiliation{ICREA-Instituci\'o
Catalana de Recerca i Estudis Avan\c{c}ats, Passeig Llu\'{\i}s Companys,
23, 08010 Barcelona, Spain}
\author{P. Jarillo-Herrero}\affiliation{Department of Physics, Massachusetts Institute of Technology, Cambridge, MA 02139, USA}
\author{F.H.L. Koppens}\email[Correspondence to: ]{frank.koppens@icfo.es} \affiliation{ICFO - Institut de Ci\'encies Fot\'oniques,
Mediterranean Technology Park, Castelldefels (Barcelona) 08860,
Spain}

\begin{abstract}
Controlling the energy flow processes and the associated energy relaxation rates of a light emitter is of high fundamental interest, and has many applications in the fields of quantum optics, photovoltaics, photodetection, biosensing and light emission. While advanced dielectric and metallic systems have been developed to tailor the interaction between an emitter and its environment, active control of the energy flow has remained challenging. Here, we demonstrate in-situ electrical control of the relaxation pathways of excited erbium ions, which emit light at the technologically relevant telecommunication wavelength of 1.5 $\mu$m. By placing the erbium at a few nanometres distance from graphene, we modify the relaxation rate by more than a factor of three, and control whether the emitter decays into either electron-hole pairs, emitted photons or graphene near-infrared plasmons, confined to $<$15 nm to the sheet. These capabilities to dictate optical energy transfer processes through electrical control of the local density of optical states constitute a new paradigm for active (quantum) photonics.
\end{abstract}

\maketitle

Spontaneous emission constitutes a canonical example of energy flow from an excited light emitter into its environment, where energy relaxation takes place via photon emission. Alternatively, for an emitter in the vicinity of a solid, energy relaxation can occur through channels involving electronic excitations, such as electron-hole pairs and collective charge oscillations (plasmons). Tailoring spontaneous emission by modifying the local density of optical states (LDOS), which governs the emitter-environment interactions \cite{Yablonovitch, Novotny2006}, has been achieved using, amongst others, optical cavities \cite{Gerard1998,Raimond2001,Englund2007,Hennessy2007}, photonic crystals \cite{Lodahl2004,Englund2005}, and metallic nanostructures \cite{Novotny2011}. In these systems the LDOS available for the light emitters is typically a fixed property that depends only on the type and geometry of the material system. Here, we control electrically and in-situ the local density of optical states and therefore the energy relaxation rate of a nearby emitter, by employing graphene. Specifically, we demonstrate in-situ tuning of the magnitude and character of the energy transfer pathways from optically excited erbium ions -- emitters for near-infrared light that are used as a gain medium in telecommunication applications \cite{Polman1997, Snoeks1995}. This control enables new avenues in a range of fields, covering photovoltaics \cite{Gratzel2001,Polman2010}, photodetection \cite{Novotny2011}, bio-sensing \cite{Anker2008}, light emission \cite{Yablonovitch, Englund2005, Achermann2004}, and active photonics \cite{Vakil2011}.
\\

The ability to control in-situ the LDOS requires a material for which the optical excitations that occur for a specific emission energy can be modified. Because graphene is gapless and it has a Fermi energy that is electrostatically tunable up to optical energies of $\sim$1 eV, it can effectively behave as a semiconductor, a dielectric, or a metal. Here, we propose to use these material characteristics to electrically control the relaxation rate and energy transfer processes of a dipolar emitter at subwavelength distance from the graphene. The concept of our experiment is shown in Fig.\ 1a, schematically representing the gate-tunable energy flow and relaxation processes, experienced by a dipolar emitter with emission energy $E_{\rm em} = 0.8$ eV at a distance of 10 nm from a graphene sheet with Fermi energy $E_F$. For $E_F < E_{\rm em}/2$, there is energy transfer from the emitter to interband electron-hole (e-h) pair excitations in graphene \cite{Swathi2008, GomezSantos2011, Velizhanin2012}. In the second regime, which crosses over at a Fermi energy of about $E_F = E_{\rm em}/2$, the emitter-graphene coupling is strongly reduced and the graphene is almost "invisible" for the emitter, with most energy of the excited emitter relaxing by photon emission.
\\

Interestingly, a third regime at $E_F  > 0.7E_{\rm em}$ is accessible, where graphene behaves as a metallic material and as a result the emitter couples to intrinsic plasmons, i.e.\ propagating electron density waves that are strongly confined to the graphene sheet (see also the Poynting vector representation in the right panel of Fig.\ 1a). Graphene plasmons have been a strong focus for active research due to their gate-tunability and the short plasmon wavelength that is $\sim$50-100 times smaller than the photon wavelength \cite{Polini2008,Jablan2009,Grigorenko2012}. However, such strongly confined graphene plasmons have only been observed in the mid-infrared (IR) \cite{Chen2012,Fei2012,Yan2013} or far-IR \cite{Yan2012,Long2011} and usually on patterned graphene for resonant plasmon excitation \cite{Long2011,Yan2013,Fang2013,Brar2013,Yan2012}. The near-field coupling between erbium emitters, which emit at 1530 nm, and strongly doped graphene makes the excitation of plasmons in the near-IR possible, since the electric field of the nearby dipolar emitters contains wave vector components that match the (relatively high) wave vector of the graphene plasmon \cite{Nikitin2011, Koppens2011}. This coupling results in the excitation of graphene plasmons confined to distances $<$15 nm from the graphene sheet.
\\

To demonstrate the in-situ tunable coupling between emitters and graphene, we use hybrid erbium--graphene devices as shown in Fig.\ 1b-c. These devices contain a thin layer ($<60$ nm) of Er$^{3+}$ emitters embedded in an oxide matrix (see Supp.\ Info) on top of a silicon substrate (Fig.\ 1b). We have chosen erbium due to its technological relevance in telecommunication applications \cite{Polman1997} and because the emission energy $E_{\rm em}$ of 0.8 eV is relatively low compared to most emitters. This enables us to access the regime where $E_F > E_{\rm em}/2$, which is required for strong modification of the energy transfer rate and for access to the plasmon regime. A Hall-bar shaped sheet of graphene is placed directly on top of the erbium--oxide layer. Thus the erbium emitters are in subwavelength proximity of the graphene, where the physics are governed by their near-field coupling. Finally, the top layer consists of a transparent solid polymer electrolyte gate that enables a gate-tunable Fermi energy above 0.8 eV \cite{Das2008}. We have studied more than ten devices of two different types and they all show very similar behavior as the results we present here. One type of device contains a thin layer of erbium-doped yttria (1-3\% atom.\ concentration) with a thickness of 45--60 nm (see Supp.\ Info) and another type contains erbium in a SiO$_2$ matrix ($<$1\% atom. concentration) with a thickness of $\sim$25 nm (see Supp.\ Info).\
We use fluorescence as a probe for the energy relaxation pathways, and a typical spatially resolved fluorescence map is shown in Fig.\ 1d. We excite erbium emitters with a 532 nm focused laser spot and measure the near-infrared erbium emission using a home-built scanning confocal microscope setup (see Supp.\ Info). The fluorescence $F_{\rm g}$  for emitters in the region with graphene is quenched by about a factor 2-3 compared to the fluorescence from erbium emitters $F_{\rm 0}$ (in the same host) without graphene on top. Similar quenching was observed earlier for visible light emitters coupled to graphene and attributed to the energy transfer from the emitters into excited e-h pairs \cite{Treossi2009,Chen2010,Gaudreau2013}. Here, graphene effectively behaves as a semi-conductor.
\\

We now show the ability to electrically control the emitter--graphene coupling, and thus the energy flow pathways, by tuning the Fermi energy of the graphene using the gate voltage of the polymer electrolyte. Figure 2a shows the fluorescence for a device containing a $\sim$25 nm thick layer of
SiO$_2$ with erbium ions, with variation of the gate voltage and laser position. The surface plot and the line cuts clearly show that around the charge-neutrality point the erbium emission in the graphene region is reduced, whereas the emission outside the graphene region is unaffected by
the gate voltage. For higher Fermi energies, $E_F>E_{\rm em}/2$, the erbium emission in the graphene region becomes similar to the erbium emission outside the graphene region. This marks the transition from the \textit{e-h pair excitation} regime to the \textit{photon emission} regime. The former regime corresponds to a strong decrease in erbium emission, whereas in the latter the graphene sheet is almost invisible for the emitter, leading to emitter fluorescence as if the graphene were not there.
\\

The key signature of electrical control of the LDOS is tunability of the relaxation rate of the erbium emitters. To demonstrate this, we perform time-resolved fluorescence measurements for a location inside the graphene region (decay rate $\Gamma_{\rm tot,g}$) and a location outside the graphene region (decay rate $\Gamma_{\rm tot,0}$) as a function of $E_F$, which are shown in Fig.\ 2b. The resulting emitter lifetimes, extracted from exponential fits to the decay curves, are shown as a function of gate voltage in Fig.\ 2c. For the energy range around the charge-neutrality point $E_F<E_{\rm em}/2$, $\Gamma_{\rm tot,g}$ is about a factor three higher than $\Gamma_{\rm tot,0}$, while $\Gamma_{\rm tot,g}$ approaches $\Gamma_{\rm tot,0}$ for $E_F>E_{\rm em}/2$. Thus for $E_F>E_{\rm em}/2$ energy relaxation occurs mainly through photon emission, whereas for $E_F<E_{\rm em}/2$ a parallel energy relaxation channel opens up, which leads to e-h pair excitation in graphene. Such strong in-situ control of the relaxation rate is a unique feature of this hybrid system.
\\

A more quantitative analysis is presented in Fig.\ 3a, where we show the experimentally obtained erbium fluorescence contrast $\eta=F_{\rm g}/F_{\rm 0} \approx \Gamma_{\rm tot,0}/\Gamma_{\rm tot,g}$ (for small pump power, see Methods) as a function of $E_F$, compared with the theoretical results that we will further discuss below. The device is the same one as in Fig.\ 1d with a $\sim$60 nm thick layer of Er$^{3+}$:Y$_2$O$_3$ (see Supp.\ Info) and the Fermi energy is calibrated through Hall measurements (see Supp.\ Info). The data and model show excellent agreement. The two theoretical curves are obtained by integrating the fluorescence of dipoles at distances between 2 and 50 nm from the graphene sheet and correspond to a rigorous simulation of the complex dielectric environment through a proper multiple-reflection formalism \cite{Blanco2004} and a simulation using a simplified dielectric environment (see also Supp.\ Info). We remark two notable features in the experimental and theoretical curves. First, the quenching in the \textit{photon emission} regime does not completely disappear. Second, the fluorescence contrast $\eta$ depends on $E_F$ in a non-monotonic fashion, and decreases for $E_F>0.6$ eV.
\\

In order to address these two intriguing features, we present here the basic elements of the model that describes the decay rate of an emitter in a simplified dielectric environment: an emitter placed near a graphene layer, supported in turn by a homogeneous substrate. These elements are qualitatively the same as in the layered structure actually used in both experiment and simulations (see Supp.\ Info). The decay rate is proportional to the electric field that is induced by a dipole on itself due to its environment \cite{Novotny2006}. The latter is well described through the Fermi-energy-dependent graphene conductivity $\sigma (E_{\rm em}, E_F)$ at fixed emission energy $E_{\rm em}$ = 0.8 eV, as well as by the substrate permittivity $\epsilon$. Because of the translational invariance of the planar surface, the self-induced field is naturally decomposed into contributions of well-defined parallel wave vector $k_{\parallel}$. The resulting rate
$\Gamma_{\rm e-g}$, normalized to the emission rate $\Gamma_0$ without
graphene, reduces to

\bea {{\Gamma_{\rm e-g}}\over{\Gamma_{\rm 0}}} - 1 \propto \int^\infty_0 dk_{\parallel} B(k_{\parallel})\Im\{r_p\} \hspace{1mm} , {\rm where} \eea

\bea r_p = {{-2}\over{\epsilon + 1 + 2 \sigma (E_{\rm em}, E_F) i k_{\parallel}\lambda_{\rm em}/c}} \hspace{1mm}  \eea

is the Fresnel reflection coefficient, and $c$ the speed of light. The
bell-shaped weight function $B(k_{\parallel})= k_{\parallel}^2 e^{-2
k_{\parallel}d}$ represents the distribution of the wave vectors that
contribute to the emitter-graphene coupling when they are separated by a
distance $d$. This distribution is peaked at $k_{\parallel}=1/d$. We take a weighted sum of the rate over emitter dipole orientations, as the latter are random. The theoretical decay rates that comprise the theoretical results as shown in Fig 3a, are shown in Fig.\ 3b as a function of Fermi energy for a few fixed distances. These traces reveal energy transfer rate increases up to a factor 3000 for $d$ = 5 nm. We note that although the electric or magnetic nature of the dipole associated with erbium emission is still a subject of investigation \cite{Li2014}, the distance and orientation averages lead to the same emission rates, when assuming the erbium emission is described by either an electric dipole or a magnetic dipole.\\

The gate-tunability of the emitter-graphene coupling is evident from Eqs.\ 1 and 2, because of the term $\sigma (E_{\rm em}, E_F)$, which makes $r_p$ depend strongly on $E_F$, as illustrated in Fig.\ 3c. The blue curve, corresponding to a relatively small $k_{\parallel}$, has the same shape as $\Re[\sigma (E_{\rm em}, E_F)]$, reflecting the excitation of electron-hole pairs through vertical transitions, which is suppressed by two orders of magnitude for $E_F>E_{\rm em}/2$. However, for larger $k_{\parallel}$, for example as represented by the red curve, the suppression is much weaker. In the experiment, these large wave vectors are present due to the very small emitter--graphene distance $d$. This explains the incomplete recovery of the fluorescence at $E_F > E_{\rm em}/2$ (e.g. in Fig.\ 3a), as illustrated in Fig.\ 1a (middle Dirac cone): even though electron-hole pair excitations with small $k_\parallel \approx 0$ (vertical transitions) are inhibited above $E_F=0.4$ eV -- larger wave vectors  still result in electron-hole pair excitations (through non-vertical transitions). We remark that this near-field effect is distinctively different from the not fully understood far-field effect of residual absorption in the Pauli-blocking regime for mid-IR light, which was attributed to a non-zero background conductivity even for large $E_F$ \cite{Basov2008}. In contrast, we can fully ascribe the observations to the incomplete recovery of the fluorescence to e-h pair excitations by higher wave vectors.
\\

Interestingly, upon further increase of the Fermi energy (above 0.6 eV), both the experimental data and the theoretical model in Fig.\ 3a-b show a decrease of emission (stronger quenching) and a reduction in lifetime. We ascribe this effect to energy transfer to near-infared graphene plasmons. Graphene plasmons emerge at larger $E_F$ because of the suppression of interband transitions and because of the increasing $\Im[\sigma (E_{\rm em}, E_F)$] with $E_F$ (i.e. graphene becomes more metallic). The plasmon response is reflected by a pole in $r_p$, as seen from inspection of Eq.\ 2. From this pole, we extract the well-known dispersion relation for the plasmon wave vector $k_{\rm sp} \approx i c (\epsilon + 1)/ (2 \sigma (E_{\rm em}, E_F) \lambda_{\rm em} )$ \cite{Wunsch2006,Hwang2007, Jablan2009,Nikitin2011, Koppens2011}. The plasmon resonance is visible in Fig.\ 3c as a narrow peak for the green line that represents large $k_\parallel$.
\\

One of the experimental signatures of coupling between the emitters and graphene plasmons is the reduction of the emitter lifetime (reduction of fluorescence), which is stronger for increasing $E_F$. This is understood by the consideration that the plasmon field, which governs the coupling strength, decays away from the graphene sheet with $e^{-k_{\perp}d}$, with $k_{\perp} \approx k_{\rm sp}$. Because $k_{\rm sp} \propto 1/E_F$, an increasing $E_F$ results in stronger plasmon coupling. This explains the downward slope of the observed $\eta$ (and thus emitter-plasmon coupling) with $E_F$.
\\

In order to further verify that the experimental observations are signatures of near-infrared graphene plasmons, and to provide more insight in the plasmon field confinement, we take advantage of the fact that the non-radiative emitter--graphene coupling decays with $d^{-4}$ \cite{GomezSantos2011,Velizhanin2012,Swathi2008,Gaudreau2013}, whereas the plasmon-emitters coupling decays exponentially: $e^{-k_{\perp}d}$ \cite{Nikitin2011,Koppens2011,Velizhanin2012}. By controlling the distance between the emitters and graphene, we can tune the relative contribution of the two coupling mechanisms and obtain an estimation of the plasmon field confinement. To this end, we use devices with an additional Al$_2$O$_3$ spacer layer of different thickness $t$ in between graphene and the
erbium layer. Half of the graphene region is in direct contact
with the erbium layer, whereas the other half is separated by the
spacer layer. In Fig.\ 4a we show the emission as a function of
Fermi energy for a region without spacer layer, with a 5 nm spacer
layer and a 12 nm spacer layer. For $E_F<E_{\rm em}/2$ (the
\textit{e-h pair excitation} regime) all curves show emission
quenching and all curves show the transition to the \textit{photon emission} regime starting at $E_F$ = 0.4 eV. The \textit{plasmon coupling} regime is clearly visible for the curve without spacer layer and the one with 5 nm spacer layer. However, the curve with 12 nm spacer layer, shows almost no plasmon coupling up to 0.8 eV. In the Supp.\ Info we show the same trends for $\sim$10 times lower excitation power, and for a sample with erbium emitters in SiO$_2$, rather than in Y$_2$O$_3$. We reproduce these trends in Fig.\ 4b, where we show the calculated emission $vs.$\ Fermi energy: with increasing spacer layer the \textit{plasmon coupling} regime starts at a higher Fermi energy. In this calculation we integrate the fluorescence from emitters at distances that range from $t$ to $t+D$, with $D$ the emitter layer thickness $D$.
\\

We can intuitively understand these observations by analysing the Fresnel coefficient $r_p$, which is shown as a function of Fermi energy and $k_{\parallel}$ in Fig.\ 4c. Here, we also include the $k_{\parallel}$-dependence of $\sigma$ through the random-phase approximation \cite{Wunsch2006, Hwang2007} (see Supp.\ Info). The bell-shaped distribution function for three different distances is shown as line traces in Fig.\ 4c. For short distances, up to $\sim$5 nm, the plasmon resonance is included for $E_F>0.6$ eV. However, for a 12 nm distance, there are no wave vectors that overlap with the plasmon resonance. For both theory and experiment, this results in a fluorescence curve that is independent of $E_F$ for $E_F>0.6$ eV. We illustrate the emitter-graphene coupling in Fig.\ 4d-f, which shows the numerically calculated electric field patterns at a distance of 5 nm, a distance of 10 nm and a distance of 15 nm. By comparing our data to the model, we can estimate $k_\perp$ of the graphene plasmons and thus the plasmon field confinement with respect to the graphene surface. We find that it is approximately 10 nm, as expected for an emission wavelength of 1.5 $\mu$m. We remark that we do not expect direct intraband excitations to be responsible for the decrease in emission for $E_F > 0.6$ eV. This can be seen in Fig.\ 4c, where it is clear that at a distance of 5 nm the wave vectors overlap with the plasmon resonance, but not with the (weaker) intraband excitations. Therefore the decreasing emission with Fermi energy is well in agreement with coupling to graphene plasmons.
\\

In conclusion, by placing a graphene sheet with controllable Fermi
energy in nanometre scale proximity of an emitter, we show
electrical tunability of its optical emission, relaxation rate and relaxation pathways. In the case where energy flow leads to e-h pair excitations or plasmons in graphene, rather than being directly lost to heat, the energy could be harvested. Furthermore, the ability to control optical fields by electric fields at length scales of just a few nanometres will open many avenues for opto-electronic nanotechnologies such as on-chip optical information processing as well as (quantum) information \cite{Yin2013} and communication schemes.

\section*{METHODS}

The total decay rate on graphene is given by $\Gamma_{\rm tot,g} = \Gamma_{\rm rad} + \Gamma_{\rm loss} + \Gamma_{\rm e-g}$, where the first term is the radiative decay rate without graphene (but with the dielectric environment), the second an intrinsic loss term in the thin emitter film, and the latter the term that describes the emitter--graphene coupling. The relation between the fluorescence inside the graphene region (subscript $g)$ and outside (subscript $0$) is given by

\bea F_{\rm g/0} \propto {{\Gamma_{\rm rad}}\over{1 +
\bigl({{\Gamma_{\rm tot,g/0}}\over{P_{\rm exc}\Gamma_{\rm exc}}} \bigr)}} \eea

Here, $\Gamma_{\rm rad}$ is the radiative decay rate for excited
erbium ions, $P_{\rm exc}$ is the excitation power, and
$\Gamma_{\rm exc}$ is the excitation rate that describes the
creation of excited state population, which we determine
experimentally (see Supp.\ Info). We also experimentally determine the loss rate $\Gamma_{\rm loss}$ from the decay rate outside the graphene region and find that the decay rate for erbium emitters in a thin oxide film is reduced with respect to erbium emitters in bulk oxide (see Supp.\ Info). This intrinsic loss mechanism is likely caused by energy scavenging of (surface) impurities and leads to the observation of a contrast of a factor 3 between emission in the graphene region and outside the graphene region, whereas this would be larger if the intrinsic loss were not present. The equation for $F_{\rm g/0}$ simplifies to $F_{\rm g}/F_{\rm 0} \approx \Gamma_{\rm tot,0}/\Gamma_{\rm tot,g}$ for small excitation power, showing that a larger decay rate directly corresponds to smaller emission. For a complete treatment, the role of the excitation power is taken into account in the model presented in Figs.\ 3 and 4.

\section*{ACKNOWLEDGEMENTS} --- KJT thanks NWO for a Rubicon fellowship. FK acknowledges support by the Fundacio Cellex Barcelona, the ERC Career integration grant 294056 (GRANOP), the ERC starting grant 307806 (CarbonLight). FK and JGdA acknowledge support by the E.C. under Graphene Flagship (contract no. CNECT-ICT-604391). JGdA acknowledges support by GRARPA. The work at MIT has been supported by AFOSR grant number FA9550-11-1-0225, a Packard Fellowship, and the MISTI-Spain program. This work made use of the Materials Research Science and Engineering Center Shared Experimental Facilities supported by the National Science Foundation (NSF) (award no. DMR-0819762) and of Harvard's Center for Nanoscale Systems, supported by the NSF (grant ECS-0335765).
\\

\onecolumngrid

\section*{FIGURES}

\begin{figure} [h!!!!!]
   \centering
  \includegraphics [scale=0.60]
   {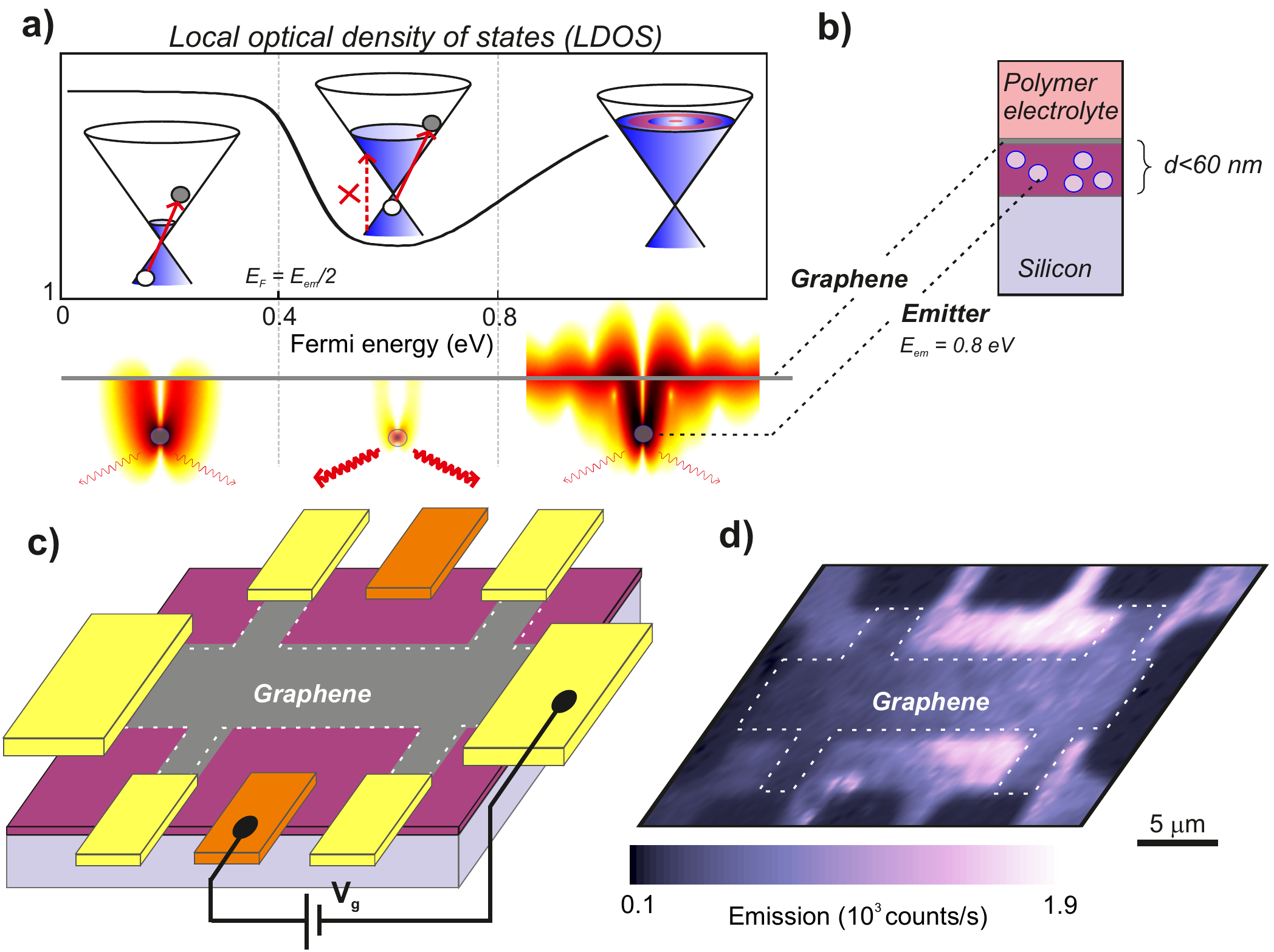}
   \caption{\textit{Concept and device for electrically controllable energy relaxation pathways.} \textbf{a)} Schematic representation of the experiment, where we tune the emitter-graphene system through three different coupling regimes by electrically controlling the Fermi energy of the graphene sheet: For low
   Fermi energy (\textit{e-h pair excitation} regime)
   the LDOS of emitters nearby graphene is increased with respect to emitters in vacuum, and efficient energy transfer occurs from the emitters to graphene, as illustrated by a color plot of the Poynting vectors for an emitter at 10 nm distance from the graphene sheet (see below the LDOS plot). Upon increasing the Fermi energy (\textit{photon emission} regime) electron-hole pair excitation is reduced (see Dirac cone) and the energy flows to emitted photons. Further increasing the Fermi energy (\textit{plasmon coupling} regime) leads to plasmons launching in graphene.
   \textbf{b)} Cross section of the device, showing the different layers.
\textbf{c)} Schematic representation of the hybrid erbium--graphene device. The device contains six
   metal contacts in a Hall bar configuration and two additional
   metal contacts that are used
   for applying a gate voltage $V_g$ to the polymer electrolyte on top of the graphene layer. \textbf{d)}
   Confocal microscopy image of the erbium emission at 1.5 $\mu$m, detected by raster scanning the device ($\sim$60 nm Er$^{3+}$:Y$_2$O$_3$) with a focused 532 nm laser beam (spot size $\sim$1 $\mu$m). This image corresponds to the \textit{e-h pair excitation} regime, where the emission is reduced and the graphene shape is clearly visible. The collected emission
   is also reduced when the metal contacts and gates are illuminated, partially due to reflection and partially due to energy transfer between erbium and the metal \cite{Novotny2006}.
}
   \label{F1}
\end{figure}

\clearpage

\begin{figure} [h!!!!!]
   \centering
   \includegraphics [scale=0.7]
   {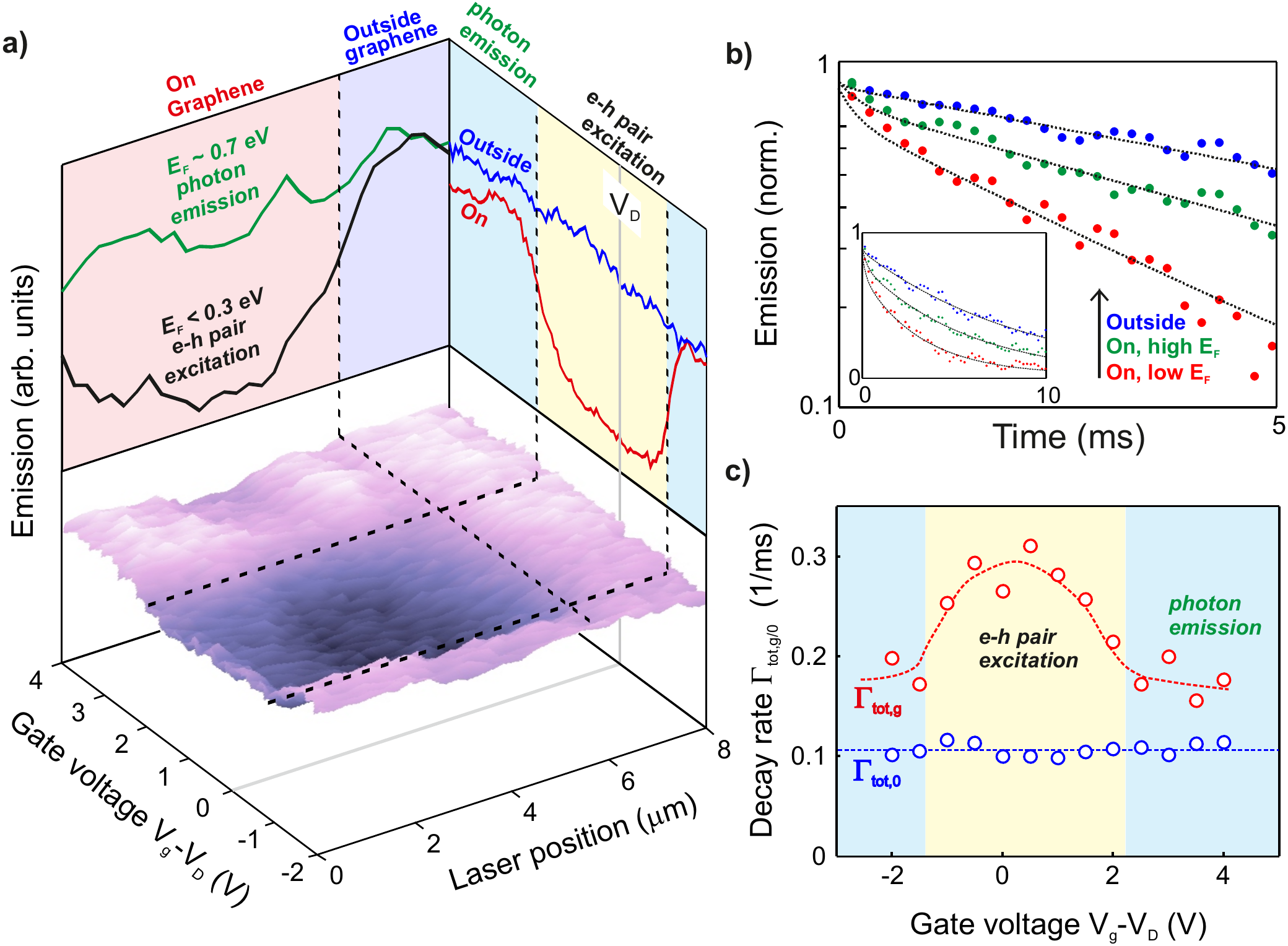}
   \caption{\textit{Electrically controlling spontaneous emission.}
 \textbf{a)} Erbium emission of a device with 25 nm Er$^{3+}$:SiO$_2$, as a function of position and gate voltage shows reduced emission at low Fermi energy and no gate voltage dependence outside graphene. The left cuts show emission \textit{vs.}\ laser position for low and high Fermi energy. The right cuts show emission \textit{vs.}\ gate voltage for laser excitation at a location inside the graphene region and a location outside the graphene region.
   \textbf{b)} Normalized emission (logarithmic scale) as a function of time after switching of a 10 ms laser pulse for positions outside and inside the graphene region. The decay rate is enhanced strongly on graphene at low $E_F$ and weakly enhanced on graphene at high $E_F$. Dashed lines are double-exponential fits, where we use the slower timescale for the decay rate $\Gamma_{\rm tot,g/0}$ for a location inside/outside the graphene region. The inset shows the time traces on a linear scale. The bi-exponential behavior is likely the result of the different erbium--graphene distances within the emitter layer.
   \textbf{c)} The extracted decay rates on graphene show strongly non-monotonous behavior as a function of gate voltage. The dashed line is to guide the eye.}
   \label{F2}
\end{figure}

\clearpage

\begin{figure} [h!!!!!]
   \centering
   \includegraphics [scale=0.7]
   {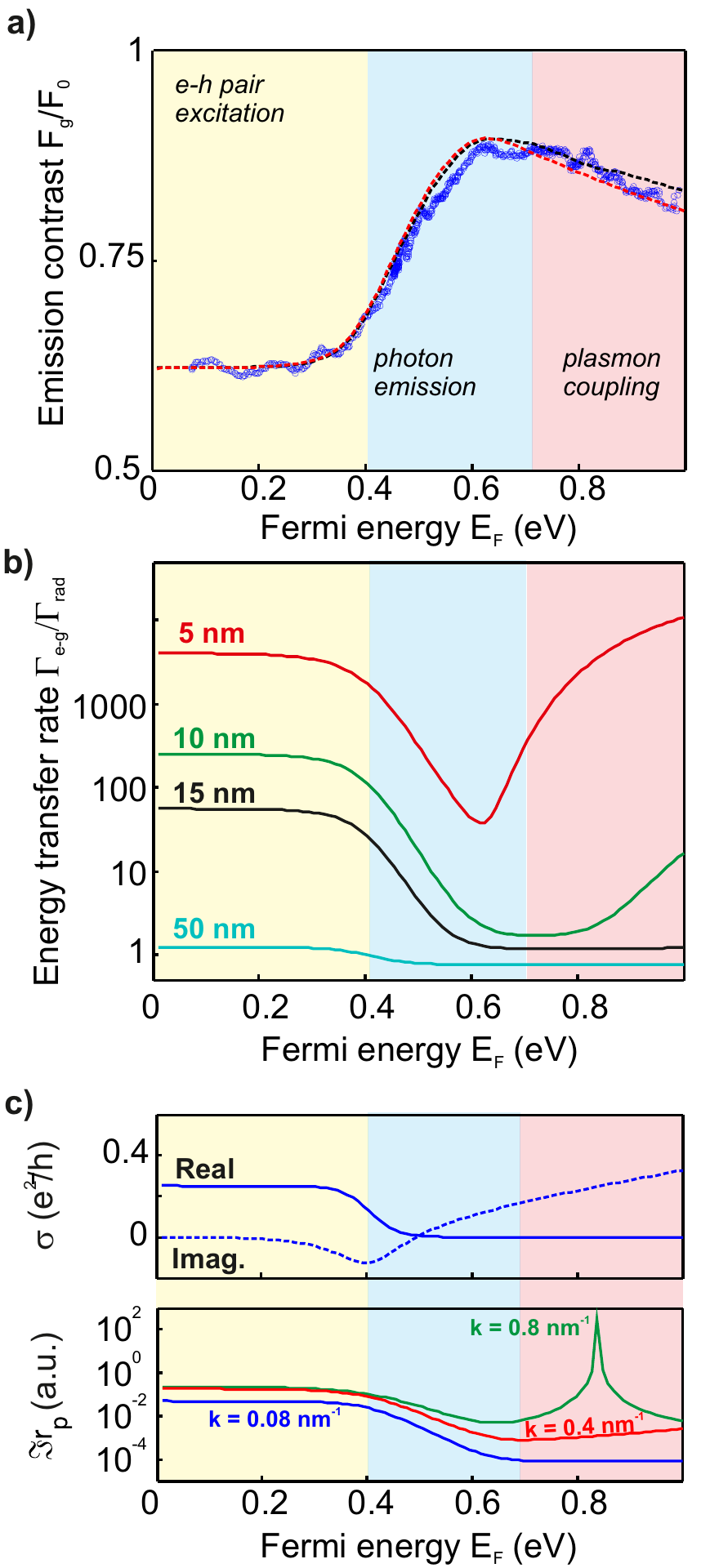}
   \caption{\textit{Comparison of experiment and theory.} \textbf{a)} Emission contrast as a function of Fermi energy for a device with $\sim$60 nm of Er$^{3+}$:Y$_2$O$_3$ together with the theoretical results considering only the silicon substrate (red dashed line) and the complete dielectric environment (black dashed line), using a mobility of 10,000 cm$^2$/Vs, a temperature of 400 K (see also Supp.\ Info), and integrating the contributions of emitters located between 2 and 50 nm from the graphene sheet. To obtain the emission contrast, we take into account the intrinsic loss in the emitter layer and the non-negligible excitation power (see Methods and Supp.\ Info). The experimental and theoretical curves show strongly reduced emission at low $E_F$, weakly reduced emission above 0.4 eV and emission decreasing with $E_F$ above 0.6 eV.
\textbf{b)} Calculated energy transfer rate that corresponds to the theoretical model (considering only the silicon substrate) in panel \textbf{a} for four different emitter-graphene distances. The shortest distance ($d = 5$ nm)
   corresponds to the largest range of wave vectors $k_{\parallel}$
   that contribute to the coupling. Therefore for this distance
   the quenching is very strong and the plasmon launching is very
   efficient. At large distances, the effects are greatly reduced.
   \textbf{c)} Top panel: complex conductivity of graphene (mobility 10,000 cm$^2$/Vs, temperature 400 K), showing the decreasing (increasing) real (imaginary) part upon increased doping. Bottom panel: the Fresnel reflection coefficient $r_p$ as a function of Fermi energy for three distinct values of $k_\parallel$, showing a decrease for $E_F > E_{\rm em}/2$ and a plasmon peak for large wave vector at a specific Fermi energy (green curve).}
   \label{F3}
\end{figure}

\clearpage

\begin{figure} [h!!!!!]
   \centering
   \includegraphics [scale=0.8]
   {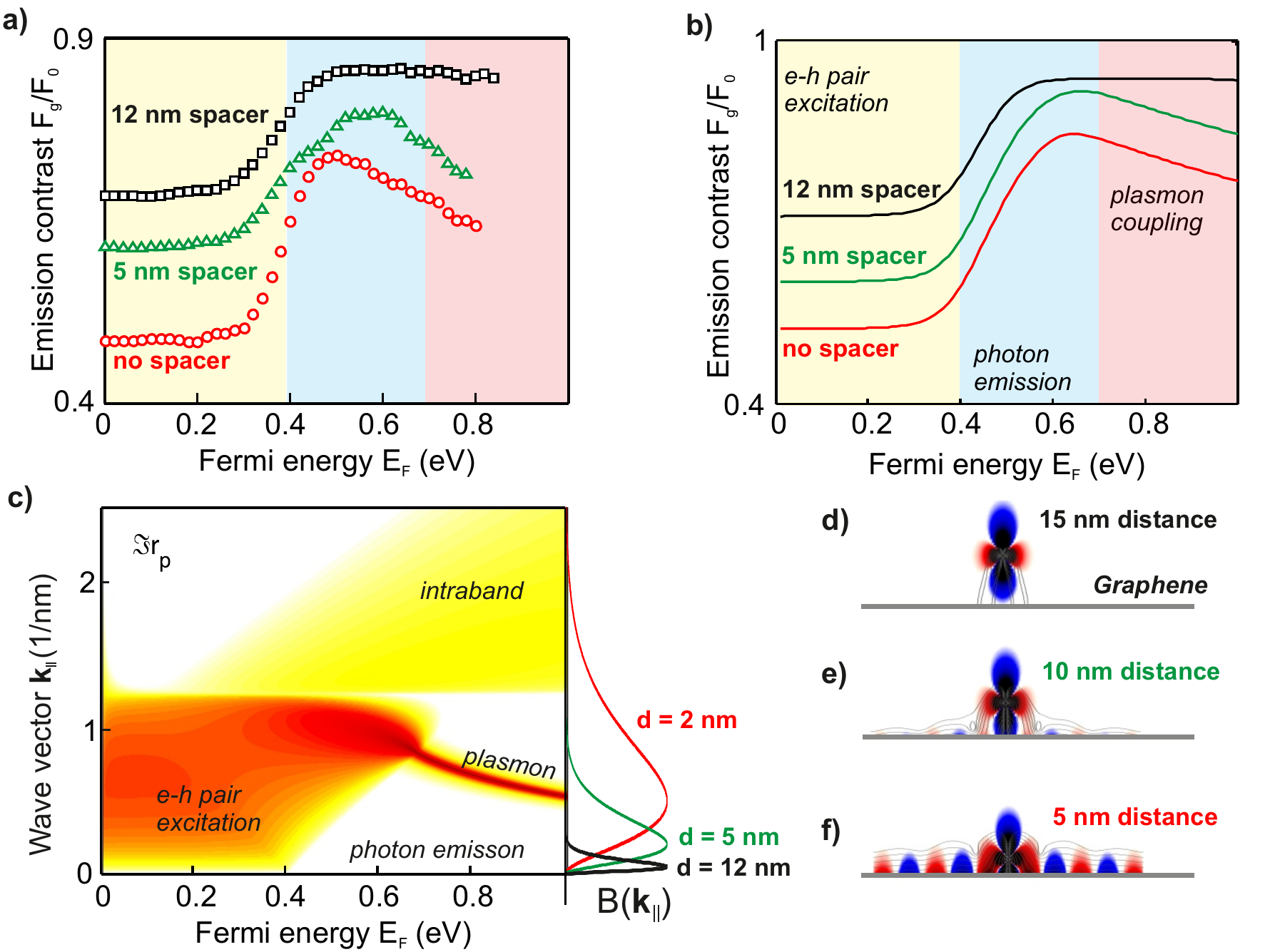}
   \caption{\textit{Strong field confinement: Plasmon launching at 1.5 $\mu$m.} \textbf{a)} The emission contrast as a function of Fermi energy for devices with a $\sim45$ nm thick Er$^{3+}$:Y$_2$O$_3$ layer separated by the graphene by an Al$_2$O$_3$ spacer layer of 5 and 12 nm. Data without spacer layer are also included.
 The \textit{plasmon coupling} regime is characterized by a downward sloping emission with Fermi energy. With a spacer layer between the emitters and graphene with thickness $t = 5$ nm the trend is very similar, although the coupling in all regimes is weaker. With a $t = 12$ nm spacer layer, there is almost no plasmon coupling visible.
\textbf{b)} Theoretical emission contrast using the same parameters as in Fig.\ 3.
\textbf{c)} Calculated $r_p$ as a function of $E_F$ and
   $k_{\parallel}$, where we now also include the $k_{\parallel}$-dependence of $\sigma$ through the random-phase approximation \cite{Wunsch2006,Hwang2007}(see Supp.\ Info). The transition from the \textit{e-h pair excitation} (red) into the \textit{photon emission} (white) regime depends on both $E_F$ and $k_{\parallel}$. Furthermore, the plasmon resonance (Eq.\ 2) is clearly visible for high Fermi energies. The plasmon resonance appears for lower wave vectors (and thus smaller graphene-emitter distances) when $E_F$ is increased. On the right side, we show the bell shape $B(k_{\parallel})$ (see Eq.\ 1) that contains all contributing wave vectors for three emitter-graphene distances $d$. \textbf{d,e,f)} Snap shots of the instantaneous electric field amplitude, demonstrating the near-field interaction between an emitter and graphene, for emitter-graphene distances of 5 nm (\textbf{d}), 10 nm (\textbf{e}) and 15 nm (\textbf{f}).}
   \label{F4}
\end{figure}

\end{document}